\documentclass{opticajnl}
\journal{pr}

\title{Spatiotemporal mode decomposition of ultrashort pulses propagating in graded-index multimode fibers}
\author[1,*]{Mario Zitelli}
\author[2]{Vincent Couderc}
\author[1]{Mario Ferraro}
\author[1]{Fabio Mangini}
\author[1]{Pedro Parra-Rivas}
\author[1]{Yifan Sun}
\author[1,3]{Stefan Wabnitz}

\affil[1]{Department of Information Engineering, Electronics and Telecommunications, Università degli Studi di Roma Sapienza, Via Eudossiana 18, 00184 Rome, Italy}
\affil[2]{Universit\'e de Limoges, XLIM, UMR CNRS 7252, 123 Avenue A. Thomas, 87060 Limoges, France}
\affil[3]{CNR-INO, Istituto Nazionale di Ottica, Via Campi Flegrei 34, 80078 Pozzuoli, Italy}
\affil[*]{Corresponding author: mario.zitelli@uniroma1.it}

\begin{abstract}
We develop a spatiotemporal mode decomposition technique to study the mode power distribution of ultrashort pulses emerging from long spans of graded-index multimode fiber, for different input laser conditions. We find that beam mode power content in the dispersive pulse propagation regime can be described by the Bose-Einstein law, as a result of the process of power diffusion from linear and nonlinear mode coupling among nondegenerate mode groups. In the soliton regime, the output mode power distribution approaches the Rayleigh-Jeans law.
\end{abstract}

\setboolean{displaycopyright}{false}

\begin{document}

\maketitle

\section{Introduction}

In recent years, there has been a resurgence of interest in studying nonlinear effects in multimode optical fibers (MMFs), both for their application and fundamental science perspectives \cite{krupa2019multimode, Cristiani2022}. Among these, the surprising observation of spatial beam self-cleaning induced by the Kerr effect in graded-index (GRIN) MMFs has attracted significant attention \cite{Krupa2017, liu2016kerr}. Analogous, but different in nature from Raman beam clean-up \cite{Chiang:92}, beam self-cleaning consists in the reshaping, as the input power is increased, of the multimode speckled transverse intensity profile at the output of the fiber into a bell-shaped beam close to the fundamental mode of the fiber, sitting on a low-intensity multimode background \cite{Krupa2017}.

Spatial beam self-cleaning can be exploited to substantially improve the beam quality at the output of MMFs, a property which is of interest for high-peak power beam delivery, mode-locked fiber lasers \cite{wright2017spatiotemporal,10.1117/1.AP.2.5.056005} and high-resolution imaging systems \cite{moussa2021spatiotemporal,Zhang:19}, to cite a few.
To date, several models have been proposed for describing the physical mechanism of Kerr self-cleaning, including nonlinear nonreciprocal mode coupling \cite{krupa2017spatial}, self-organized instabilities \cite{Wright2016}, 2D hydrodynamic turbulence \cite{Podivilov2019}, beam condensation \cite{aschieri2011condensation,PhysRevLett.125.244101} and, last but not least, beam thermalization \cite{wu2019thermodynamic}. In a thermodynamic framework, the multimode beam is represented as a gas of indistinguishable particles, that populates the large, but finite number of spatial modes of the fiber. This approach is powerful, since it permits to exploit the conservation laws (e.g., the number of particles, or power, the linear momentum and the orbital angular momentum), which are inherent to virtually lossless beam propagation in short lengths of fiber. 
By exploiting the tools of statistical mechanics, one may then describe beam self-cleaning as the irreversible evolution of the gas of photons towards its state of thermal equilibrium. This is identified by the definition of thermodynamic parameters, such as the temperature ($T$) and the chemical potential ($\mu$) \cite{wu2019thermodynamic}. The validity of this approach has been recently confirmed by means of mode-decomposition experiments, revealing the input power dependence of the mode redistribution at the fiber output, in good qualitative agreement with theoretical predictions \cite{pourbeyram2022direct,mangini2022statistical,podivilov2022thermalization}.   

In the thermodynamic approach, the role of nonlinearity is merely that of inducing, via four-wave mixing, the exchange of photons among nondegenerate fiber modes, in analogy with particle collisions in a gas \cite{wu2019thermodynamic}. However, it has been noted that mode energy exchanges leading to thermal equilibrium are greatly facilitated by the presence of random linear mode coupling (RMC)
\cite{PhysRevLett.122.123902,SIDELNIKOV2019101994}. The question then arises, is nonlinearity a necessary ingredient for thermalization in highly MMFs, or linear disorder can do the job by itself? And in this case, what is the expected output mode power distribution? In Raman fiber laser experiments based on km-long spans of GRIN MMF, it has been observed that the highly multimode pump beam reaches a steady-state distribution that corresponds to mode equipartition \cite{Kuznetsov:21,Kuznetsov2021sr,Kharenko:22}.
Whereas in a recent work, the roles of nonlinear and linear mode coupling have been systematically studied, by artificially varying the strength of linear disorder in a MMF \cite{PhysRevLett.129.063901}. These experiments confirmed that linear and nonlinear coupling processes may lead to different thermalization regimes, which are consistent with either the mode equipartition (when linear disorder prevails) or a Rayleigh-Jeans (RJ) distribution (when nonlinearity dominates).
In \cite{PhysRevLett.129.063901}, strong linear disorder was introduced by applying a stress to the MMF via a series of clamps. It remains to be studied what is the role of the linear disorder which is inherent to standard telecommunication GRIN fibers, owing to the presence of microbends that lead to nearest-neighbour coupling of power among adjacent nondegenerate modes \cite{Olshansky:75}, and its interplay with a weak fiber nonlinearity. 

This is the purpose of this work: by applying a novel time-resolved mode decomposition technique, which permits to measure with unprecedented accuracy the energy fraction of higher-order modes, we could determine the evolution of the mode power distribution in long lengths (up to 5 km) $L$ of GRIN MMF. We explored four propagation regimes: (i) a \textit{weakly nonlinear} dispersive pulse regime, where the effective interaction length $L_e$ was of the same order as the nonlinear length, as in the early self-cleaning experiments \cite{krupa2017spatial}, but with $L_e\ll L$; (ii) a \textit{pseudo-linear}  pulsed regime \cite{Zitelli:9306846}, where strong dispersion-induced pulse broadening rapidly imposes linear propagation, (iii) a purely linear continuous-wave (CW) regime, and (iv) a nonlinear solitonic propagation regime.

In the former three cases, linear random mode coupling prevails over weakly nonlinear mode mixing in determining the steady-state mode power distribution, which is a realistic case in most applications. Surprisingly, our observations reveal that, even in these regimes, no mode equipartition was observed in GRIN fibers. To the contrary, the long-distance mode distribution converges towards a thermalized state, that is quantitatively well described by an exponential decay vs. mode energy, as described by a 
Bose-Einstein (BE) distribution. Note that a similar exponential mode power distribution was recently observed for the Stokes beam in a Raman fiber laser; however, in that case mode-dependent gain has a significant influence in determining the equilibrium mode distribution \cite{Kharenko:22}. 
In the soliton case, beam condensation in the fundamental fiber mode and pulse narrowing were observed. Here the mode power distribution approaches the Rayleigh-Jeans law, albeit with a slightly larger depletion of higher-order modes. Our experiments revealed the occurrence of a sudden change of the output modal distribution, from the BE to a quasi-RJ law, when a temporal soliton is forming. This finding provides, we believe for the first time, an experimental evidence of a distribution transition, which occurs in long spans of multimode optical fiber when passing from a linear-disorder dominated and pseudo-linear dispersive propagation regime, to a soliton regime.


\section{Theory}

Modes of GRIN fibers are defined in terms of radial and azimuthal numbers ($\ell,m$), and can be grouped in terms of their quantum number $q = 2\ell + |m|+1$, with $q=1, 2, .., Q$ 
because of their equal propagation constant, $\beta_q$. Besides the total number of power packets $N=\sum n_{q}$, the total momentum or energy $H = \sum \epsilon_{q} n_{q}$, where $\epsilon_q= \beta_q-\beta_Q$,
is also conserved upon beam propagation.
Maximization of the entropy, constrained by the conservation of $N$ and $H$, leads to the BE law \cite{wu2019thermodynamic,mangini2022statistical}

\begin{equation}
|c_q|^2=\frac{n_q}{g_q}=
\frac{1}{\exp[-(\mu+\epsilon_q)/T]-1}\label{eq:BE}
\end{equation}
where $g_q=q$ is the number of modes within group $q$. 
Whenever $|(\mu+\epsilon_q)/T|\ll 1$, Eq.~(\ref{eq:BE}) can be approximated by the Rayleigh-Jeans (RJ) distribution $|c_q|^2=
-T/(\mu+\epsilon_q)$.
\noindent As we shall see, for high-order modes (i.e., with small $\epsilon_q$) the  approximation leading to the RJ law becomes invalid.
Meaning that the experimental mode distribution is described by the original BE law Eq.~(\ref{eq:BE}). In the limit of high temperatures, the mode distribution can be approximated with the Maxwell-Boltzmann (MB, or exponential) law $|c_q|^2\simeq\exp[(\mu+\epsilon_q)/T]$, i.e., the equilibrium distribution of a gas of distinguishable classical particles.
In the experiments described in the next section, mode group powers $n_q$ were estimated from fast photodiode traces. The average individual mode power fraction was determined as $|c_i|^2=n_q/q$, with $i=1$ for $q=1$, $i=2, 3$ for $q=2$,  $i=4, 5, 6$ for $q=3$, and so on. This is justified by considering that RMC randomly scrambles the power of degenerate modes; hence, a statistical equipartition of mode power within each group can be assumed.
Distributions of $|c_i|^2$ for the individual modes will be compared with BE, RJ, and MB laws, limiting the analysis to $Q=10$ groups comprising $N=55$ modes. 

In order to verify the accuracy of a mode decomposition, it is necessary to perform a reconstruction of the output near-field intensity pattern. 
When using 2D modal decomposition methods such as in Ref.\cite{Flamm:12}, it is challenging to properly reconstruct the output near-field of ultrashort pulses propagating over long fiber lengths, because of the temporal variation of the mode envelope phases. Such variations are due to: (i) the chromatic dispersion induced phase chirp of pulses carried by different modes; (ii) the mode dispersion induced group delay among modes; (iii) laser-induced phase noise. 

To overcome these difficulties, it is necessary to adopt a 3D mode field reconstruction method which accounts for all of the previously listed effects, which enables us to reconstruct both the near-field transverse intensity pattern and the temporal power traces, emerging from long spans of GRIN fibers. Our 3D field reconstruction method uses the experimental output power fractions $|c_i|^2$, in order to provide a weight to the numerically propagated output spatiotemporal mode fields. 
We simulated the linear propagation, in the absence of 
fiber nonlinearity and mode coupling, of each individual mode field. This provides the estimated time-dependent output mode phase profile, which accounts for the cumulated chromatic and modal dispersion, and laser linewidth; the relative phase among different modes was randomly chosen, as expected from linear RMC. We suppose that each mode carries a Gaussian pulse 
with initial complex amplitude $A_p(0,t)$ and 3D field

\begin{equation}
\begin{aligned}
	E_i(x,y,z=0,t)=A_i(0,t)F_i(x,y)=\sqrt{P_i}exp\left[i\phi_i\right]exp\left[-\frac{(t-t_0)^2}{2T_0^2} \right]F_i(x,y),
	\label{eq:ModeBullet}
	\end{aligned}
\end{equation}

\noindent where $F_i(x,y)$ is the mode eigenfunction, $P_i$ is the input peak power, $T_{FWHM}=1.665T_0$ is the constant pulse width, and $\phi_i$ is an input random phase, modeling the effect of linear RMC during propagation.
In order to simulate the CW beam propagation regime, the Gaussian pulse was replaced by a laser noise term $exp\left[i\int_{-\infty}^{t}\omega_0(t')dt'\right]$, where the laser linewidth noise $\omega_0(t)$ has a zero mean and standard deviation $\sigma_{\omega}=2\pi B$, with $B$ the laser linewidth. 

Pulse envelopes $A_i(z,t)$ were numerically propagated 
according to the different modal and chromatic dispersion, and with constant linear loss \cite{Poletti:08}. The resulting output spatiotemporal field can be written as $E_{tot}(x,y,z,t)=\sum_{i=1}^{N}c_iE_i(x,y,z,t)$.
By integrating $|E_{tot}|^2$ vs. time, one may obtain the output near-field. By integrating it vs. $(x,y)$, the instantaneous power profile can also be reconstructed.

\section{Experimental Results}\label{sec:Experiment}

In our experiments, we used relatively long (up to 5 km) spans of commercial GRIN OM4 fiber, with 50/125 $\mu$m core/cladding diameter, cladding index $n_{clad}$ = 1.444 at 1550 nm, and relative index difference $\Delta$ = 0.0103. The chromatic dispersion was $\beta_2$=-28.8 ps$^2$/km for the fundamental mode at 1550 nm, with a dispersion slope $\beta_3$=0.153 ps$^3$/km. The corresponding self-imaging (SI) period $\Lambda=\pi a/\sqrt{2\Delta}=0.55$ mm, where $a$ is the core radius.
Fibers were spooled on a support with 8 cm radius of curvature; care was taken to avoid stronger bending. According to Gloge theory \cite{Gloge:72}, bending losses remain negligible up to the first 10 mode groups ($\alpha \simeq 8\text{x}10^{-10}$ dB/km for group 10, and $\alpha >> 30$  dB/km for group 11).



Several coherent sources were alternatively used at the fiber input: a 70 fs pulse source at 1550 nm wavelength, 100 kHz repetition rate, obtained from an optical parametric amplifier (OPA) fed by a femtosecond Yb laser; a 1.4 ps unchirped pulse source, obtained from the 70 fs pulses by narrow band-pass filtering; a CW DFB laser with 1 nm linewidth at 1460 nm. The input beam was attenuated, linearly polarized and passed through a $\lambda /4$ waveplate, in order to generate a circular state of polarization. The Gaussian shaped beam was then injected into GRIN fiber spans of variable lengths (from 1 m up to 5 km), with a waist $w_0$ of approximately 15 $\mu$m, nearly twice the fundamental mode waist. The corresponding SI-induced beam compression factor $C=2\Lambda/(\pi\beta_0 w_0^2)=0.262$, where $\beta_0=2\pi n_0/\lambda$ and $n_0$ is the core index \cite{Karlsson:92,Ahsan:19}; the effective beam waist for determining the nonlinear coefficient is $w_e=\sqrt{C}w_0= 7.7\mu$m \cite{Ahsan:18,Ahsan:19}, close to the fundamental mode waist. 
The circular state of polarization was used at the input, in order to minimize power exchanges among polarization modes.
Input and output states-of-polarization were checked by measuring the power after a rotating linear polarizer, both after the input waveplate and after 5 km of GRIN fiber; the power was constant at any angle of the polarizer, with a tolerance of 5\%.
In some tests, a diffuser was introduced at the input, in order to increase the input high-order mode content. Alternatively, the input of the fiber was laterally shifted with respect to the beam by up to $\pm15$ $\mu$m, again to increase the contribution from higher-order modes. 

At the fiber output, a micro-lens focused the near-field on an InGaAs camera (Hamamatsu C12741-03); a second lens focused the beam into a real-time multiple octave spectrum analyzer (Fastlite Mozza) with a spectral detection range of 1100-5000 nm. The output pulse temporal shape was inspected by a fast photodiode (Alphalas UPD-35-IR2-D) and a real-time oscilloscope (Teledyne Lecroy WavePro 804HD) with 30 ps overall response time. An intensity autocorrelator (APE pulseCheck 50) with femtosecond resolution was also used for the characterization of the input pulses. An $M^2$ tester (Gentec Beamage M2) was used to characterize the output beam quality. Input and output power were measured by a power meter with $\mu$W resolution.



\begin{figure}[h]
\includegraphics[width=0.6\textwidth]{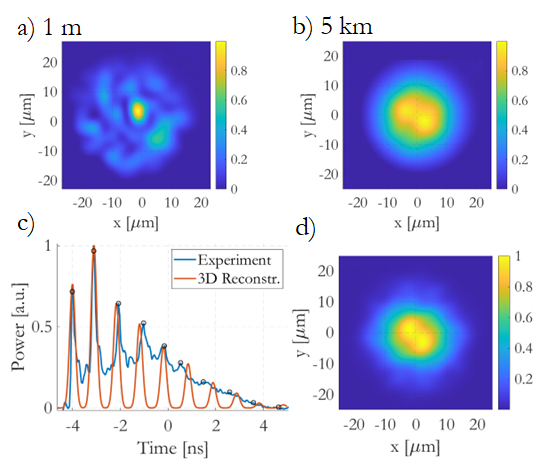}	\centering	
\caption{ 
Experiments with 1 m and 5 km of GRIN MMF, 1.4 ps pulses at 1550 nm, 100 pJ pulse energy, lateral shift of 4 $\mu$m. a) Measured near-field at 1 m. b) Measured near-field at 5 km. c) Instantaneous power at 5 km (measured and reconstructed). d) Reconstructed near-field at 5 km.
}
\label{fig:Fig1}
\end{figure}

Fig.\ref{fig:Fig1} shows the results of a first test, performed with unchirped 1.4 ps pulses at 1550 nm, propagating over 5 km of GRIN fiber. No diffuser was used at the input, and the beam lateral shift was limited to 4 $\mu$m. In this case, fiber chromatic dispersion broadens the pulse up to 285 ps, whereas modal dispersion separates the mode groups by up to 9 ns. The total pulse energy in the fiber was 100 pJ, which corresponds to $P_0=67$ W of input pulse peak power. By considering the power decrease due to dispersive pulse spreading and the weak uniform linear loss, the effective nonlinear interaction length is $L_e=144$ m; the corresponding phase shift induced by the self-phase modulation (SPM) is $\Phi_{NL}=\gamma P_0L_e=5.51$ rad, where $\gamma=2\pi n_2/(\lambda A_{eff})= 0.57  (Wkm)^{-1}$ and $A_{eff}=\pi w_e^2$. Here we used the nonlinear coefficient of silica $n_2=2.7\times 10^{-20} m^2/W$. This means that the experiment was performed with significant nonlinearity at the fiber input ($L_{NL}\simeq L_e$); nonlinear effects can be neglected after the effective distance $L_e=144$ m, and up to the fiber length of 5 km.


The measured near-field after 1 m of fiber, shown in Fig.\ref{fig:Fig1}a, is characterized by a highly speckled intensity pattern, denoting the presence of a large number of modes. On the other hand, after 5 km of propagation (Fig.\ref{fig:Fig1}b), a much brighter (nearly bell-shaped) beam is observed. The output power measured after 5 km is illustrated in Fig.\ref{fig:Fig1}c, blue curve, showing a train of temporally broadened pulses, corresponding to the propagated mode groups (up to 10 among them are visible). 

\begin{figure}[h]
\includegraphics[width=0.6\textwidth]{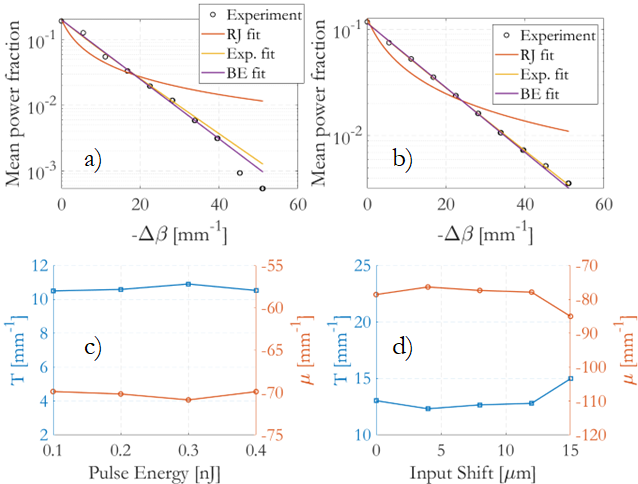}	\centering	
\caption{ 
a) Mean power fractions for test with 1.4 ps, 5 km, 100 pJ, shift 4 $\mu$m. b) Test with 70 fs, 830 m, 100 pJ, shift 15 $\mu$m. c) $T$ and $\mu$ for the test of case a) vs. pulse energy. d) $T$ and $\mu$ for the test of case b) vs. lateral shift.
}
\label{fig:Fig3}
\end{figure}

A mode decomposition (MD) for the mode powers was obtained by directly measuring the total degenerate mode group power from the photodiode response, and dividing it by the number of modes in that group; in this way one readily obtains the mean power fraction of individual modes in a degenerate group $|c_i|^2$, as shown in Fig\ref{fig:Fig3}a (black circles) against the differential propagation constant of the mode $-\Delta\beta_i=\beta_1-\beta_i$ (mm$^{-1}$). 

Fig.\ref{fig:Fig1}d shows the near field after 5 km of pulse propagation, reconstructed by using the 3D method, based on the measured $|c_i|^2$ values from Fig.\ref{fig:Fig1}c, and setting random input mode phases $0<\phi_i<2\pi$: the image cross-correlation \cite{Manuylovich:2020} between Figs. \ref{fig:Fig1}b and \ref{fig:Fig1}d is 96\%. The effect of input random phases results in a smoothing of the output near-field; the images cross-correlation and the near-field shape are modestly affected by a change of the input random mode phase sequence. 
Fig. \ref{fig:Fig1}c (orange curve) is the 3D reconstruction of the output optical power, showing a good agreement with the experiment.

\begin{figure}[h]
\includegraphics[width=0.6\textwidth]{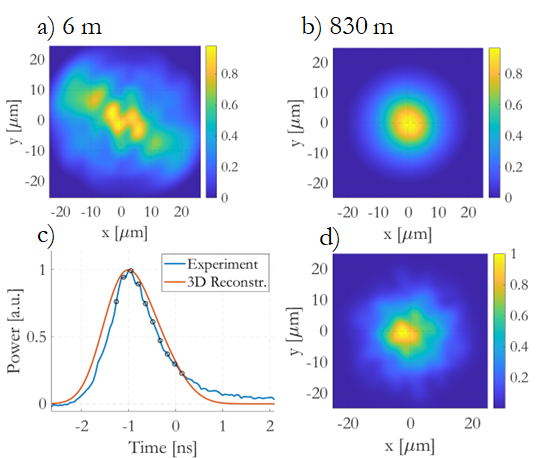}	\centering	
\caption{ 
Experiment with 6 m and 830 m of GRIN MMF, 70 fs pulses at 1550 nm, 100 pJ pulse energy, and lateral input beam offset of 15 $\mu$m. a) Measured near-field at 6 m. b) Measured near-field at 830 m. c) Instantaneous power at 830 m (measured and reconstructed). d) Reconstructed near-field at 830 m.
}
\label{fig:Fig2}
\end{figure}

In a second test, unchirped 70 fs pulses at 1550 nm were propagated over 830 m of GRIN fiber; no diffuser was used at the input, and the beam lateral shift was extended up to 15 $\mu$m with respect to the fiber axis. In this case the input pulse energy was limited to 100 pJ, giving a peak power of 1.3 kW. Because of the strong dispersive spreading, the effective nonlinear interaction length is of only $L_e=0.64$ m, and the overall nonlinear phase shift $\Phi_{NL}=\gamma P_0L_e=0.49$ rads; hence, $L_{NL}> L_e$, and the propagation regime of this test can be considered as quasi-linear, or pseudo-linear \cite{Zitelli:9306846}.

Fig.\ref{fig:Fig2}a shows the measured near-field after 6 m of fiber, which reflects the highly multimodal distribution that is injected at the fiber input. After 830 m of propagation, the measured near-field of Fig.\ref{fig:Fig2}b  shows an homogeneous, bell-shaped output beam. The experimental instantaneous power after 830 m in Fig.\ref{fig:Fig2}c, blue curve, shows a 1100 ps pulse with a trailing tail. Here chromatic dispersion is responsible for pulse broadening up to 947 ps, whereas modal dispersion introduces a 0.19 ps/m delay among adjacent mode groups. Experimentally, the overall dispersion introduces a total pulse broadening of up to 1100 ps. As a result, the different pulses belonging to nondegenerate mode groups remain strongly overlapped in time at the fiber output. Yet, it is still possible to recover the mode power distribution by sampling the measured instantaneous power at the appropriate modal delays for each mode group. After dividing these values by the number of modes in each group, one obtains the mode power distribution of Fig.\ref{fig:Fig3}b (black circles). 
Again, the measured mode power distribution permits, in combination with the 3D simulation method 
and by setting random mode phases, to reconstruct the transverse near-field intensity profile after 830 m of GRIN fiber, as shown in Fig.\ref{fig:Fig2}d. The corresponding simulated temporal power profile is reported as an orange curve in Fig.\ref{fig:Fig2}c. The good reconstruction in both spatial and temporal dimensions confirms the correct sampling of the mode group powers in Fig.\ref{fig:Fig2}c

Figs.\ref{fig:Fig3}a and \ref{fig:Fig3}b show that the modal mean power fractions can be properly fitted by exponential curves. Recall that experiments in Fig.\ref{fig:Fig3}a were conducted in a propagation regime which was initially dominated by nonlinearity, followed by a much longer propagation distance where linear disorder completely prevails. Whereas experiments in Fig.\ref{fig:Fig3}b were only weakly affected by nonlinearity even initially. In Figs.\ref{fig:Fig3}a,b we compare experimental results against different theoretical distributions. Specifically, the RJ law, the exponential MB law, 
and the BE law of Eq.~(\ref{eq:BE}). In both cases, the BE distribution better approximates the experimental data up to the 9-th mode group. Whereas the RJ law is not suitable for describing the output mode distribution, since it fails to fit the data after the 5-th mode group. The MB exponential distribution also fits well the data.
By considering the BE law as the appropriate distribution, we computed the modal temperature $T$ and chemical potential $\mu$ (mm$^{-1}$) for the first experiment (1.4 ps pulse, 5 km) when pulse energy is increased from 100 pJ to 400 pJ, see Fig.\ref{fig:Fig3}c.
As can be seen, $T$ and $\mu$ remain nearly constant when increasing the nonlinearity, as it is expected from theory \cite{wu2019thermodynamic}. 

For the second experiment (70 fs over 830 m in pseudo-linear regime), Fig.\ref{fig:Fig3}b illustrates the mean power fractions when the input beam is coupled with 15 $\mu$m lateral shift with respect to the fiber axis: as can be seen, also in this case the BE law fits very well the data. Fig.\ref{fig:Fig3}d shows that, when varying the input lateral shift from 0 to 15 $\mu$m, $T$ and $\mu$ remain nearly unchanged up to the limit value of 15 $\mu$m; this indicates that the equilibrium output mode distribution is nearly independent of the input coupling conditions.

Finally, we underline that, at variance with the approximation of mode degeneration, in our experiments we found that the values of $\Delta\beta_i$ are slightly different between the modes within the same group. This difference was accounted for when using Eq.~(\ref{eq:BE}) for fitting the experimental data in Fig.\ref{fig:Fig3}.

\begin{figure}[h]
\includegraphics[width=0.7\textwidth]{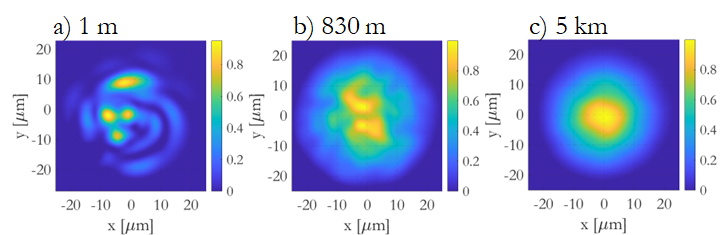}  \centering	
\caption{ 
Experiments with 1 m, 830 m, and 5 km of GRIN fiber, with input CW beam at 1460 nm, 1 $\mu$W power, an input beam diffuser and a lateral shift of 15 $\mu$m. } 
\label{fig:Fig4}
\end{figure}

In order to test a case where beam propagation occurs in the strictly linear regime, we used a CW laser source at 1460 nm. Figure \ref{fig:Fig4} shows examples of measured near-field intensity profiles after 1 m, 830 m and 5 km of GRIN fiber, respectively. At the input, a beam diffuser was added; moreover, the beam center was nominally laterally offset by 15 $\mu$m with respect to the fiber axis. The CW input power was of only 1 $\mu$W, corresponding to $L_{NL}=3\times 10^6$ km. Meaning that in this case propagation occurs in the purely linear regime: RMC is the only source of disorder. As can be seen, the near-field of Fig.\ref{fig:Fig4}a at 1 m is highly multimodal; for increasing distances, linear disorder leads to a progressive beam smoothing, producing the output of Fig.\ref{fig:Fig4}c after 5 km. 
Because of the operation in the CW regime, it was not possible to temporally resolve the mode groups in this case, in order to estimate the output mode power distribution. Still, the generation of bright beam with a near-field intensity profile similar to the case of Fig.\ref{fig:Fig2}b was observed even in this case. 

\begin{figure}[h]
\includegraphics[width=0.7\textwidth]{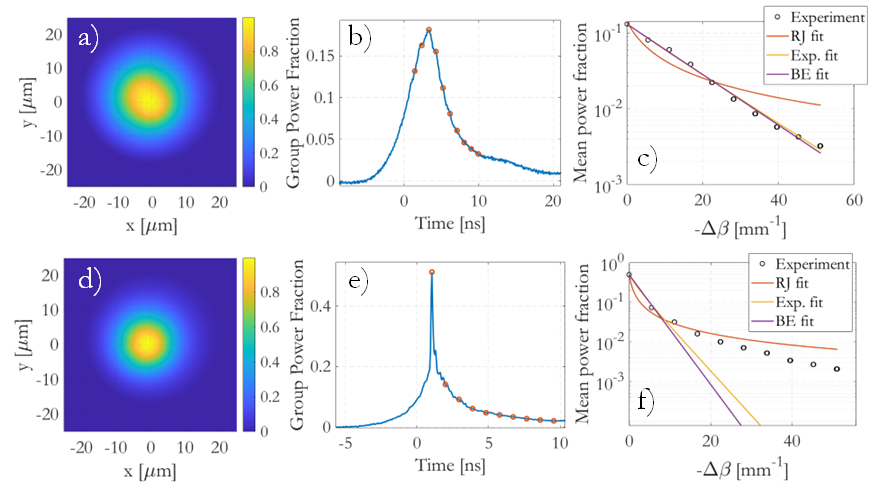}	\centering	
\caption{ 
Experiments with 5 km of GRIN MMF, 70 fs pulses at 1550 nm, lateral shift of 10 $\mu$m. a-c) Measured near-field, output power, and mean modal power fraction for input pulse energy of 200 pJ. d-f) Measured near-field, output power, and mean modal power fraction for input pulse energy of 1.3 nJ, corresponding to the formation of a multimode soliton.
}
\label{fig:Fig5}
\end{figure}

Previous experiments involved either highly dispersive pulses, of the CW propagation regime. In the former cases, the influence of the temporal dimension on the final output power distribution can be simply taken into account by considering an effective nonlinear length. This is no longer the case when the input pulse powers increased up to the point of forming multimode GRIN solitons \cite{Ahsan:18,Zitelli2021,Zitelli2021a}. To explore the soliton regime and the resulting output mode power distribution, input 70 fs pulses with energies ranging from 0.2 nJ to 1.5 nJ were propagated over 5 km of GRIN fiber. The input beam was coupled with 10 $\mu$m lateral shift with respect to the fiber axis. At 1.3 nJ of input pulse energy, the pulse starts forming a multimode soliton \cite{Zitelli2021a}.
We observed that, for input pulse powers slightly below the soliton value, the output beam was spatially compressed, while its temporal duration broadened to a few hundred of femtoseconds. Yet, a Raman-shifted walk-off soliton did not emerge; we will refer to this state as quasi-soliton regime. As soon as the walk-off soliton energy was reached (around 1.5 nJ at 1550 nm, not shown), nearly all of the output pulse energy emerged in the fundamental fiber mode. 

Figures \ref{fig:Fig5} compares the dispersive pulse with the quasi-soliton pulse regime. In the latter, Fig.\ref{fig:Fig5}e, there is no Raman-induced pulse delay, and a soliton has not yet separated from its residual dispersive waves; hence, a time-resolved modal decomposition is still possible. Figs.\ref{fig:Fig5}a-c show the measured near field, the output power (with sampled mode group powers) and the mean mode power fractions, respectively, in the dispersive pulse propagation regime. An output beam waist of 13.5 $\mu$m and a beam quality factor $M^2=1.75$ was measured. In this case, the average mode power fractions, that we have extracted from the photodiode trace, are again properly fitted by a BE law. However, when propagating in the multimode quasi-soliton regime, Figs.\ref{fig:Fig5}d-f shows the emergence of a condensed beam, with a waist of 9.4 $\mu$m and a nearly diffraction-limited beam quality factor $M^2=1.05$. This shows that most of the soliton beam power is carried by the fundamental mode. Therefore, it is reasonable to sample the first modal group in correspondence with the pulse peak in the photodiode trace. Now, Fig.\ref{fig:Fig5}f reveals that the corresponding mode power distribution no longer follows a BE law, but rather a RJ distribution. Although, the MD experiments show that higher-order modes are slightly more depleted than the prediction based on the RJ law. We may thus conjecture that the transition from the dispersive to the soliton pulse regime is akin to a phase transition, which separates qualitatively different thermalization regimes.

\section{Discussion and conclusions}

In this work, a time-resolved MD technique was used, in order to analyze the output mode power distributions after long spans of GRIN fibers. We considered four different propagation regimes, i.e., purely linear, pseudo-linear, dispersive nonlinear and soliton regimes. Our 3D MD and reconstruction method accounts for phase chirping, modal delay and random complex mode scrambling, induced by  propagation to the complex amplitude of the spatiotemporal field. The accuracy of the method was confirmed by successfully reconstructing the output field, both in the spatial and in the temporal domains. Comparisons among the experimental MD based on fast photodiode traces, the 3D modal reconstructions, and fitting of the output modal content, have shown that the BE law is the most appropriate for describing long-distance multimode fiber propagation for both dispersive nonlinear and quasi-linear regimes, in the presence of random mode coupling. Similar results in terms of near-field transverse intensity profiles were obtained in a purely linear propagation regime, although no direct confirmation of the output mode power distribution was possible in this case. 

It was also observed 
that the multimode soliton regime is characterized by a different mode power distribution,
which approaches the RJ law.
The BE distribution, observed in a regime where linear disorder prevails over a weak nonlinearity, reveals the presence of a high-temperature thermalization process, where the input power is diffused out of the fundamental mode into higher-order modes.
To the contrary, spatiotemporal beam condensation is approached in the soliton regime, and the output mode power distribution approaches the RJ law in this case, in agreement with former studies of Kerr-driven spatial beam self-cleaning.

We underline that the exponential decay of the mode power fraction must not be confused with the classical Maxwell-Boltzmann law. Indeed, in the latter, the gas particles are supposed to be distinguishable. Whereas, our experiments involve indistinguishable photons in the fiber modes, whose thermalization process occurs before the mode distinguishability provided by the modal walk-off.

\begin{backmatter}

\bmsection{Funding} This work was supported by: the European Research Council (740355, 101081871), Marie Sklodowska-Curie Actions (101064614,101023717), and Ministero dell'Istruzione, dell'Università e della Ricerca (R18SPB8227).

\bmsection{Acknowledgements} We acknowledge helpful discussions with G. Steinmeyer about using either the RJ or the BE law for describing the mode power distribution of high-order fiber modes.

\bmsection{Disclosures} The authors declare no conflicts of interest.


\smallskip

\end{backmatter}

\bibliography{References}

\bibliographyfullrefs{References}

\end{document}